\documentclass[12pt,a4paper]{article}

\usepackage{indentfirst}
\usepackage{times}
\usepackage[font=small,labelfont=bf]{caption} 
\usepackage{amsfonts, amsmath, amsthm, amssymb} 
\usepackage{wrapfig} 
\usepackage{graphicx} 
\usepackage[utf8]{inputenc}
\usepackage[english]{babel}
\usepackage{setspace}
\usepackage{authblk}
\usepackage{color}
\usepackage[symbol*]{footmisc}
\usepackage{url} 
\usepackage{wrapfig}
\usepackage{cite}
\usepackage{mdframed}

\graphicspath{{}} 

\begin{document}
\sloppy
\noindent\textbf{\Large Josephson oscillation linewidth of ion-irradiated YBa$_2$Cu$_3$O$_7$  junctions}\par 
\vspace{10pt}
\noindent A.~Sharafiev$^{1,2}$\footnote[1]{aleksei.sharafiev@espci.fr}, M.~Malnou$^1$, C.~Feuillet-Palma$^1$, C.~Ulysse$^3$, P.~Febvre$^4$, J.~Lesueur$^1$, N.~Bergeal$^1$\par
\vspace{7pt}
\noindent\scriptsize$^1$Laboratoire de Physique et d'Etude des Mat\'eriaux - UMR8213-CNRS-ESPCI Paris-UPMC, PSL Research University,10 Rue Vauquelin - 75005 Paris, France.\\$^2$Skobeltsyn Institute of Nuclear Physics, Lomonosov Moscow State University, Moscow, Russia\\$^3$Laboratoire de Photonique et de Nanostructures LPN-CNRS, Route de Nozay, 91460 Marcoussis, France.\\$^4$IMEP-LAHC - UMR 5130 CNRS, Universit\'e Savoie Mont Blanc, 73376 Le Bourget du Lac cedex, France.\par

\renewcommand{\thefootnote}{\arabic{footnote}}

\vspace{10pt}
\noindent\small\textbf{Abstract.} We report on the noise properties of ion-irradiated YBa$_2$Cu$_3$O$_7$ Josephson junctions. This work aims at investigating the linewidth of the Josephson oscillation with a detector response experiment at $\simeq$132 GHz. Experimental results are compared with a simple  analytical model based on the Likharev-Semenov equation and the de Gennes dirty limit approximation. We show that the main source of low-frequency fluctuations in these junctions is the broadband Johnson noise and that the excess ($\frac{1}{f}$) noise contribution does not prevail in the temperature range of interest, as  reported in some other types of high-T$_c$ superconducting Josephson junctions. Finally, we discuss the interest of ion-irradiated junctions to implement frequency-tunable oscillators consisting of synchronized arrays of Josephson junctions.  
\normalsize
\section{Introduction}
Ion-irradiating technology for high temperature superconducting (HTS) integrated circuits is a very promising approach for a broad range of applications because of the possibility to design circuits that include a high density of arbitrarily located Josephson elements. Developments in the field include the fabrication of SQUIDs \cite{tinchev1993a,bergeal2006}, large arrays of Josephson junctions \cite{cybart2009,cybart2014}, SQIFs \cite{ouanani2016}, heterodyne mixers for  THz frequency range \cite{malnou2012,malnou2014}, and  the recent realization of SIS tunnel junctions \cite{cybart2015}. Today, one of the main challenges in the field is to build a THz frequency-tunable oscillator by synchronizing the oscillations of a large number of Josephson junctions.  Indeed, when phase locking of $N$ junctions is achieved, the maximum output power of the array increases as $N^2$, enabling the generation of sizable power \cite{barbara1999}. In addition, the linewidth $\Delta f$ of the radiation can be considerably reduced since it decreases as $\frac{1}{N}$ \cite{darula1999}. This is particularly interesting in the case of practical HTS junctions, which have a rather broad Josephson oscillation linewidth at the typical temperature of operation ($\sim$ GHz for T$\sim$ 50~K). Recently, c-axis mesa of the layered high-temperature superconductor Bi$_2$Sr$_2$CaCu$_2$O$_{8+\delta}$ has emerged as a promising coherent THz oscillator. Its operation relies on the synchronization of the natural intrinsic Josephson junctions that are formed between CuO$_2$ superconducting planes separated by BiSrO  insulating planes \cite{welp2013}. Engineering arrays of reproducible nanofabricated HTS junctions should enable the realization of similar oscillators with an additional degree of freedom regarding the architecture of the array and possibly the frequency tunability.\par
Unlike in common types of self-excited oscillators, the high-frequency linewidth $\Delta f$ of the Josephson oscillation is mainly determined by low-frequency noise level \cite{likharev1986dynamics}. This makes any prediction difficult since low-frequency noise depends not only on well understood thermal fluctuations but also on excess $\frac{1}{f}$ noise (critical current fluctuations), whose nature is still not fully understood and which varies significantly from one type of junctions to another \cite{kawasaki1992,kemen1999,hao1996,anton2012,hts_electr_handbook}. This results in unpredictable linewidth behavior with temperature - some junctions demonstrate purely temperature defined noise level \cite{divin1992} while for others, generation linewidth might even drop down with temperature \cite{divin1993}.  The aim of this work is to measure the linewidth of ion-irradiated Josephson junctions using the self-detected dc-response technique which was previously exploited for Josephson spectroscopy and oscillation linewidth measurements of several types of junctions \cite{divin1992,divin1993,tarasov1999} and to compare the results with analytical predictions for a pure thermal fluctuation source.
\section{Experiment}
\begin{figure}[!t]
\includegraphics[width=0.4\linewidth]{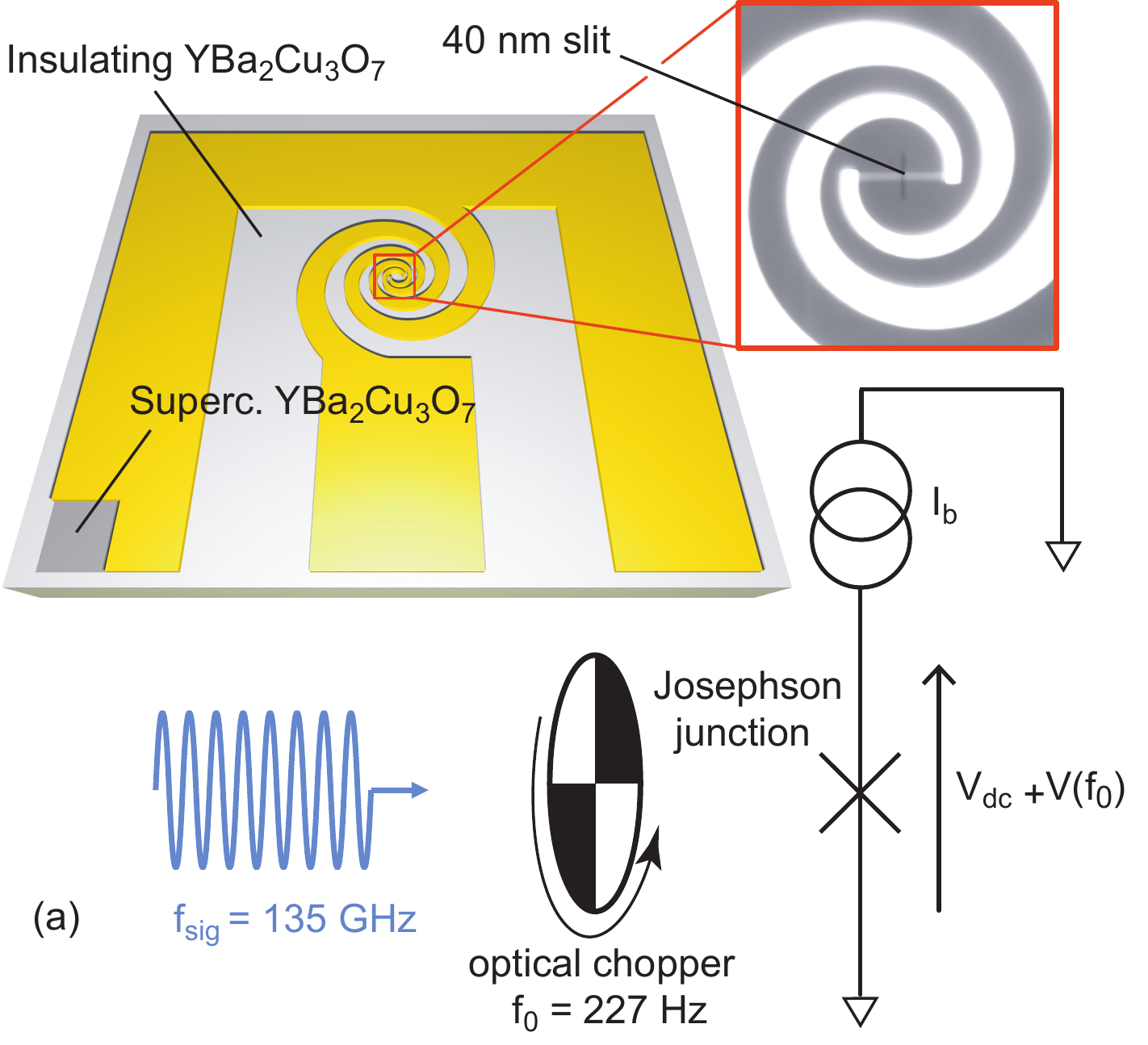}\hspace{1cm}
\includegraphics[width=0.5\linewidth]{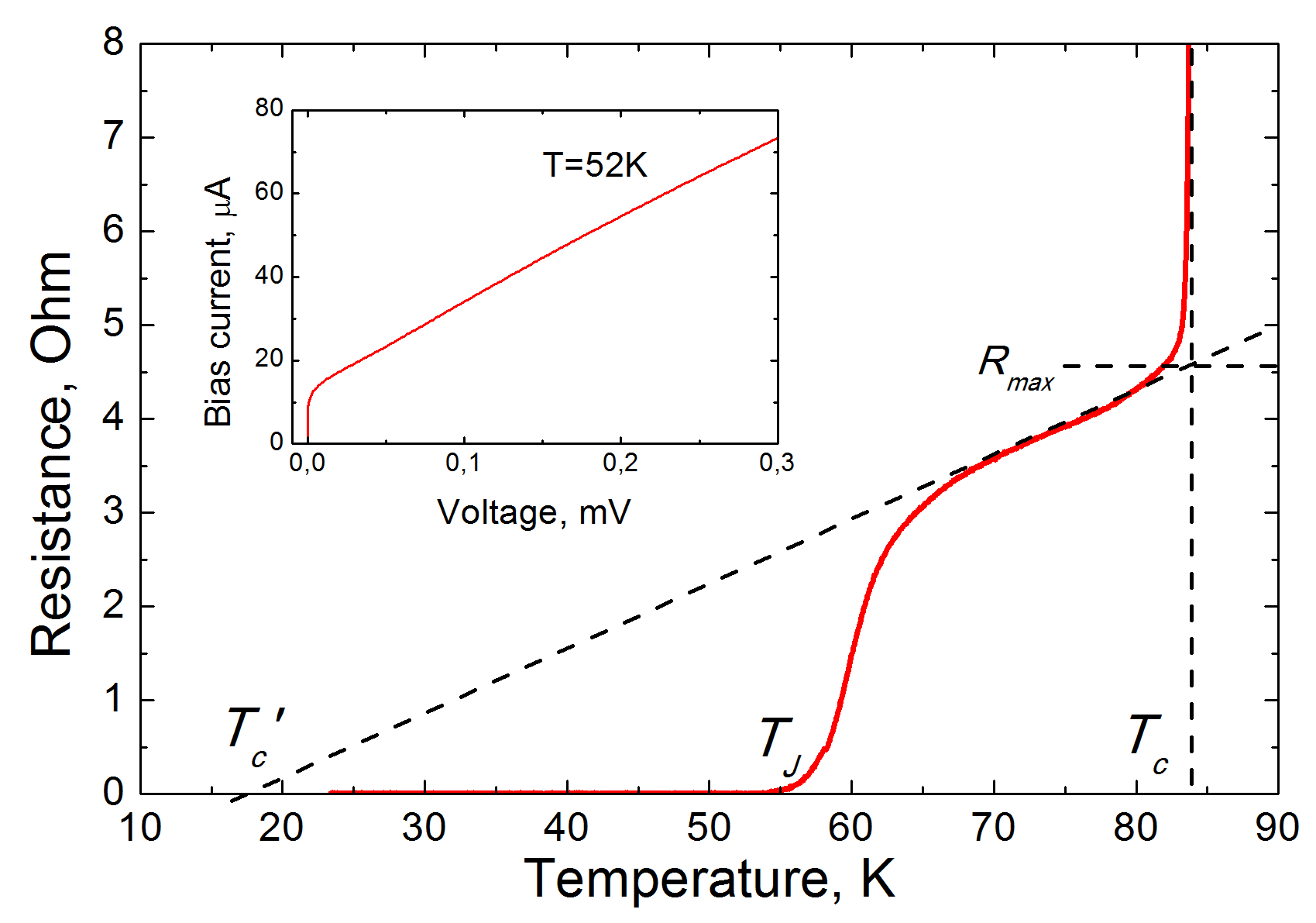}\par
\hspace{4.3cm}\textit{(a)}\hspace{8.2cm}\textit{(b)}
\captionof{figure}{\textit{(a)} Schematic description of the sample used in the detector response experiment. \textit{(b)} Experimental dependence of the junction resistance with temperature (solid red lines). An intersection of a dashed extrapolation line with X-axis can be considered as an estimation of $T_{c}^\prime$ - transition temperature for the damaged area of the junction. Inset: IV-curve of the junction for $T$=52K.}  
\label{fig1}
\vspace{-0.5cm}
\end{figure}

In this experiment we used ion-irradiated  YBa$_2$Cu$_3$O$_7$ Josephson junctions embedded into a spiral log-periodic antenna and a 50\thinspace$\Omega$ coplanar waveguide (CPW) transmission line (figure \ref{fig1}(\textit{a})). Details of the fabrication process can be found in references \cite{bergeal2007, malnou2014}. In short, we start from a commercial  70~nm thick YBa$_2$Cu$_3$O$_7$ films grown on a Al$_2$O$_3$ substrate  \cite{theva}, and covered with a 200~nm gold layer. The spiral antenna and the CPW transmission line are first defined in the gold layer through a MAN e-beam  patterned resist,  followed by a 500-eV Ar$^+$ Ion Beam Etching. Then, a 0.75-2~$\mu$m wide channel located at the center of the antenna is patterned in a MAN e-beam resist, followed by a 70-keV oxygen ion irradiation at a dose of $2\cdot 10^{15}$\thinspace at/cm$^2$.  This process ensures that the regions of the film which are not protected either by the resist or by the gold layer become deeply insulating. Finally, the junction is defined at the center of the superconducting channel by irradiating through a 40-nm wide slit patterned in a PMMA resist with 110\thinspace keV oxygen ions at  a dose of $3\cdot 10^{13}$\thinspace at/cm$^2$. The results presented in the paper were obtained with a $L$=0.75~$\mu$m-wide junction, though all measured samples demonstrated qualitatively a similar behavior.\par
\begin{figure}[!b]
\includegraphics[width=0.5\linewidth]{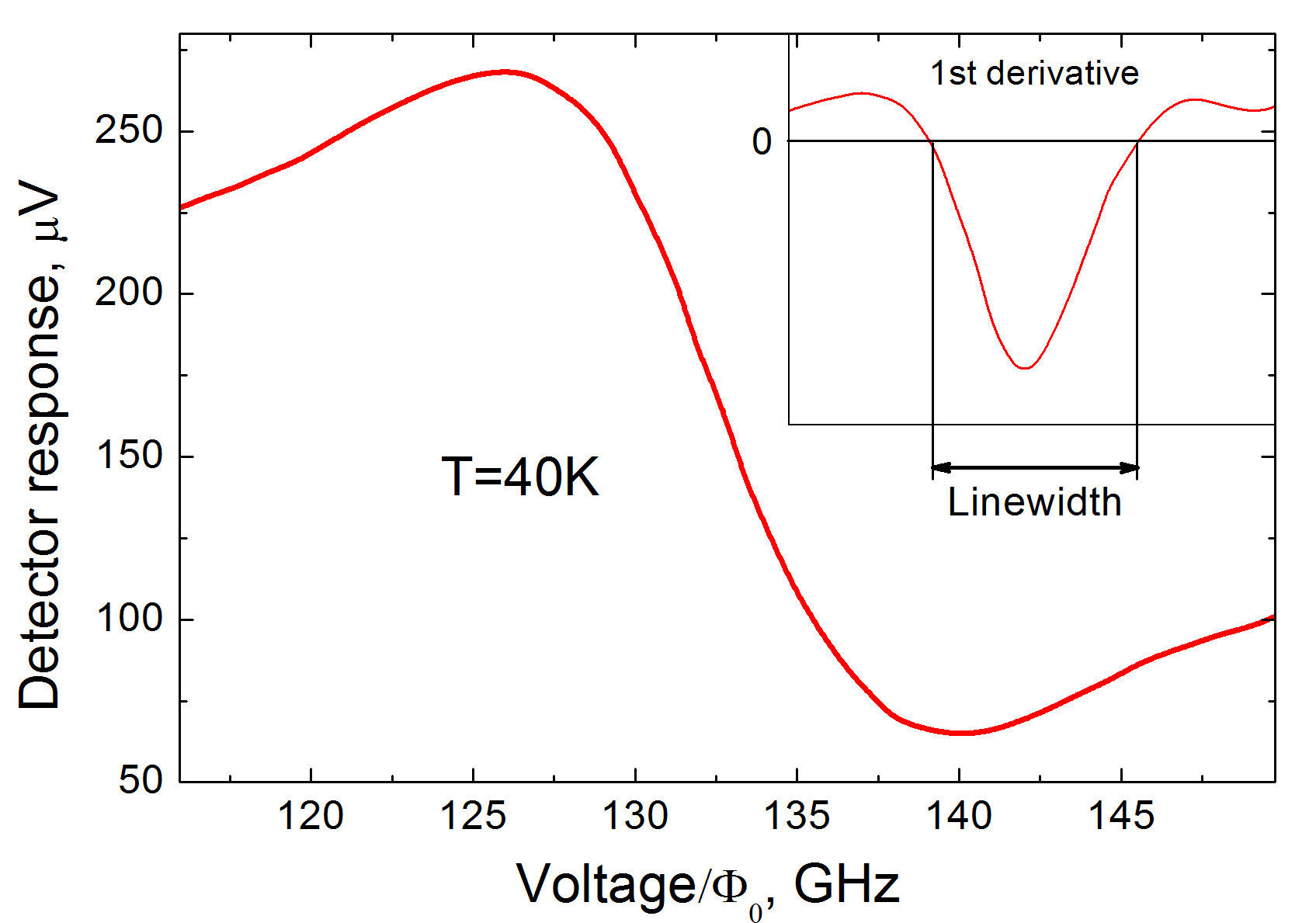}\hspace{1cm}
\includegraphics[width=0.4\linewidth]{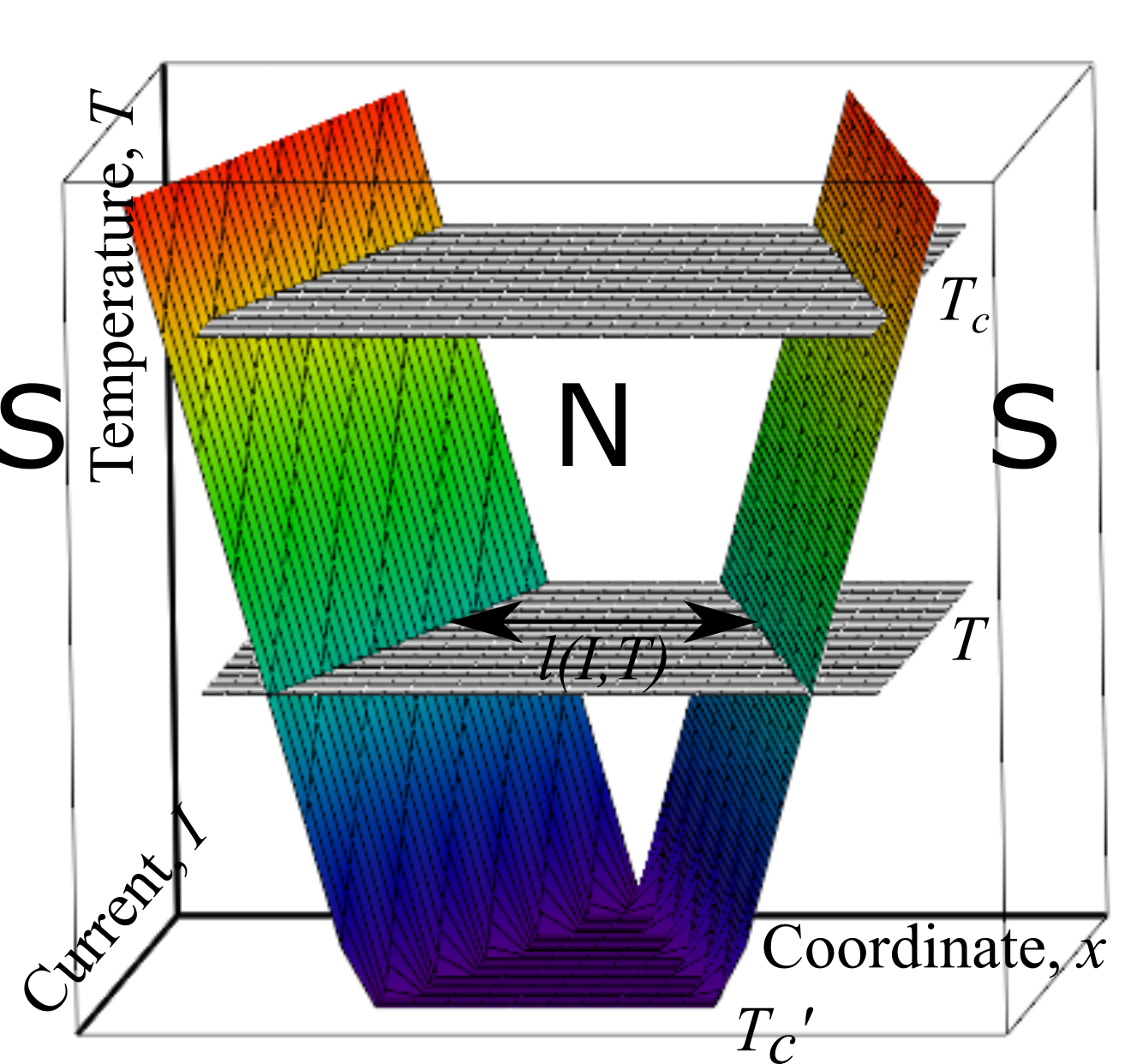}\par
\hspace{4.3cm}\textit{(a)}\hspace{8.2cm}\textit{(b)}
\captionof{figure}{(\textit{a}) Experimental detector response for temperature $T$=40K, $\Phi_{0}$ is the magnetic flux quantum.  The Josephson generation linewidth is given by the  distance between the two extrema. Inset shows the first derivative of the response. (\textit{b})  Linear dependence of the critical temperature of the ion-irradiated YBa$_2$Cu$_3$O$_7$ with bias current and coordinate assumed in the model.}
\label{fig2}
\vspace{-0.5cm}
\end{figure}
Figure \ref{fig1}(\textit{b}) shows the resistance of the junction as a function of temperature. The highest transition at T$_{c}$= 84 K refers to the superconducting transition of the reservoirs, i. e. the non-irradiated part of the YBa$_2$Cu$_3$O$_7$  channel, whereas the second transition at the lower temperature T$_J$= 55K corresponds to the occurrence of a sizable Josephson coupling (i.e stronger than thermal fluctuations, \cite{bergeal2005}). The temperature $T_{c}^{\prime}\simeq$20 K indicates the superconducting transition of the irradiated part itself and separates the pure Josephson regime from a flux-flow regime. Junctions fabricated by this method have non-hysteretic current-voltage characteristics expected in the overdamped regime, defined by a McCumber dimensionless parameter much smaller than one (inset figure \ref{fig1}(\textit{b})).\par
In order to measure the Josephson oscillation linewidth, we used the indirect detector response technique proposed in reference \cite{divin1980}. The schematic diagram of the experiment is shown in  figure \ref{fig1}(\textit{a}). A Gunn diode generates a continuous f$_{sig}$=132~GHz signal which is mechanically modulated by an optical chopper at a frequency f$_0$=227~Hz. The sample was placed inside a closed cycle refrigerator, at the focal point of a Winston cone, and was exposed to the external signal through  an optical window. The output voltage $V(f_0)$ was measured as a function of the dc bias with a lock-in amplifier synchronized at the chopper frequency $f_0$. Figure \ref{fig2}(\textit{a}) shows a typical result obtained in the Josephson regime at T=40K. The linewidth is directly extracted by measuring the spectral width between the two response extrema or  equivalently  between the zeros of its first derivative. Its evolution with temperature is shown in figure \ref{fig3}(\textit{b}). A clear broadening of the linewidth is observed when the temperature is increased which suggests a mechanism based on thermal noise consideration. In the next section, we discuss the dependencies of the critical current and thermal linewidth broadening of the junction on temperature in the framework of the de Gennes proximity effect model and the Resistively Shunted Junction (RSJ) model.  

\section{Models and discussions}
In the following, we consider the junction as a Superconductor-Normal-Superconductor type to discuss its  transport properties and generation linewidth. Notice that the normal part of the junction is itself superconducting with a reduced critical temperature $T_{c}^{\prime}$, a situation often referred as a SS'S junction. A characteristic trait of ion-irradiated junctions is  their non pure RSJ-like current-voltages curves (\cite{Katz2000,cybart_thesis}), which can be associated with temperature and bias current dependent length of the normal segment (and correspondingly normal resistance). Following this approach, we use a modified RSJ-model  and the de Gennes proximity effect model \cite{de_gennes_1964} to analyze the experimental results. The approach is illustrated in figure \ref{fig2}(\textit{b}) which shows the dependence of the YBa$_2$Cu$_3$O$_7$  critical temperature on both the coordinate $x$ along the junction and the bias current $I$. For the sake of simplicity we consider the temperature and the bias current as independent variables neglecting therefore heating effect.\par    
\subsection{Critical current }
\begin{figure}[!t]
\includegraphics[width=0.5\linewidth]{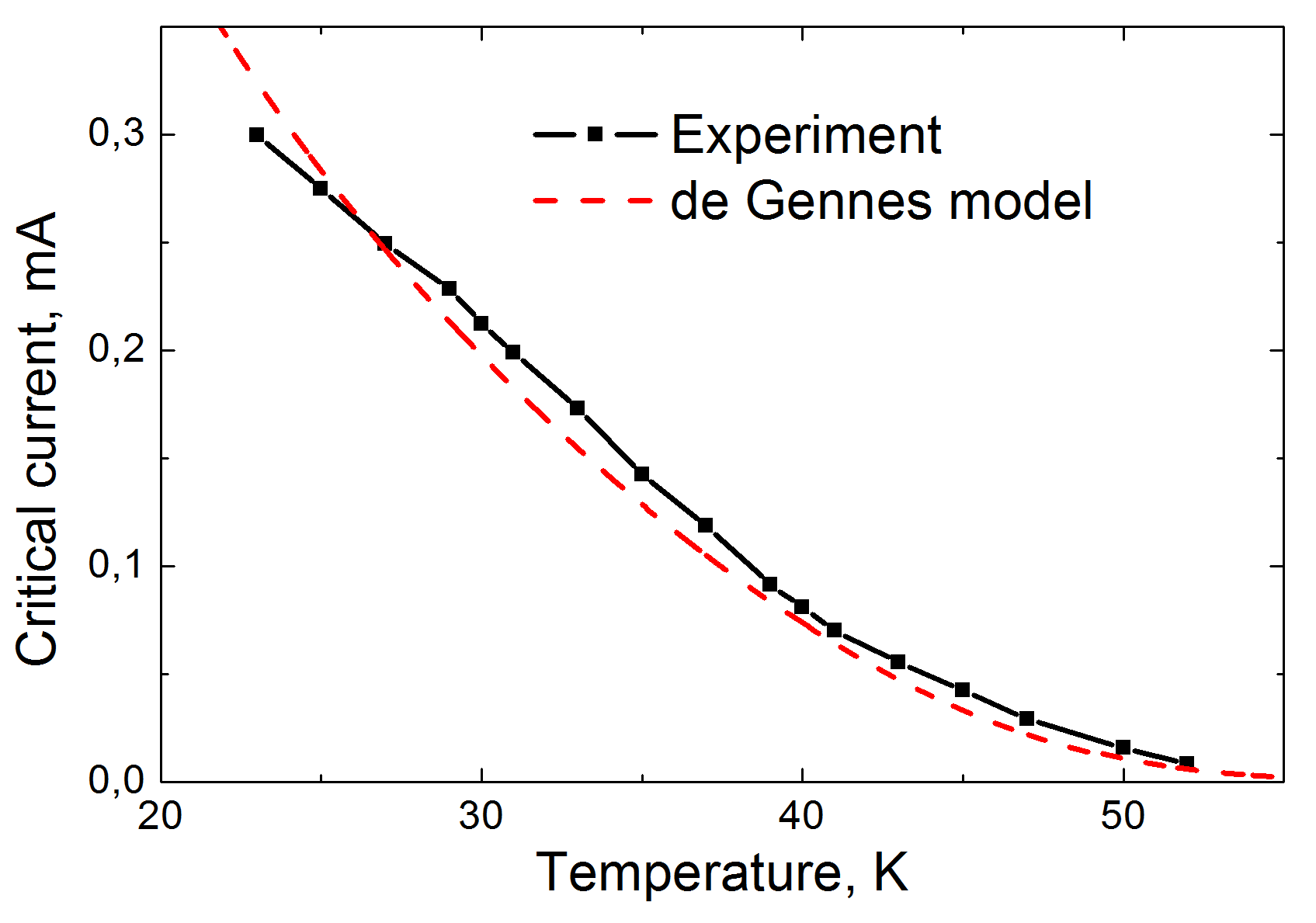}
\includegraphics[width=0.5\linewidth]{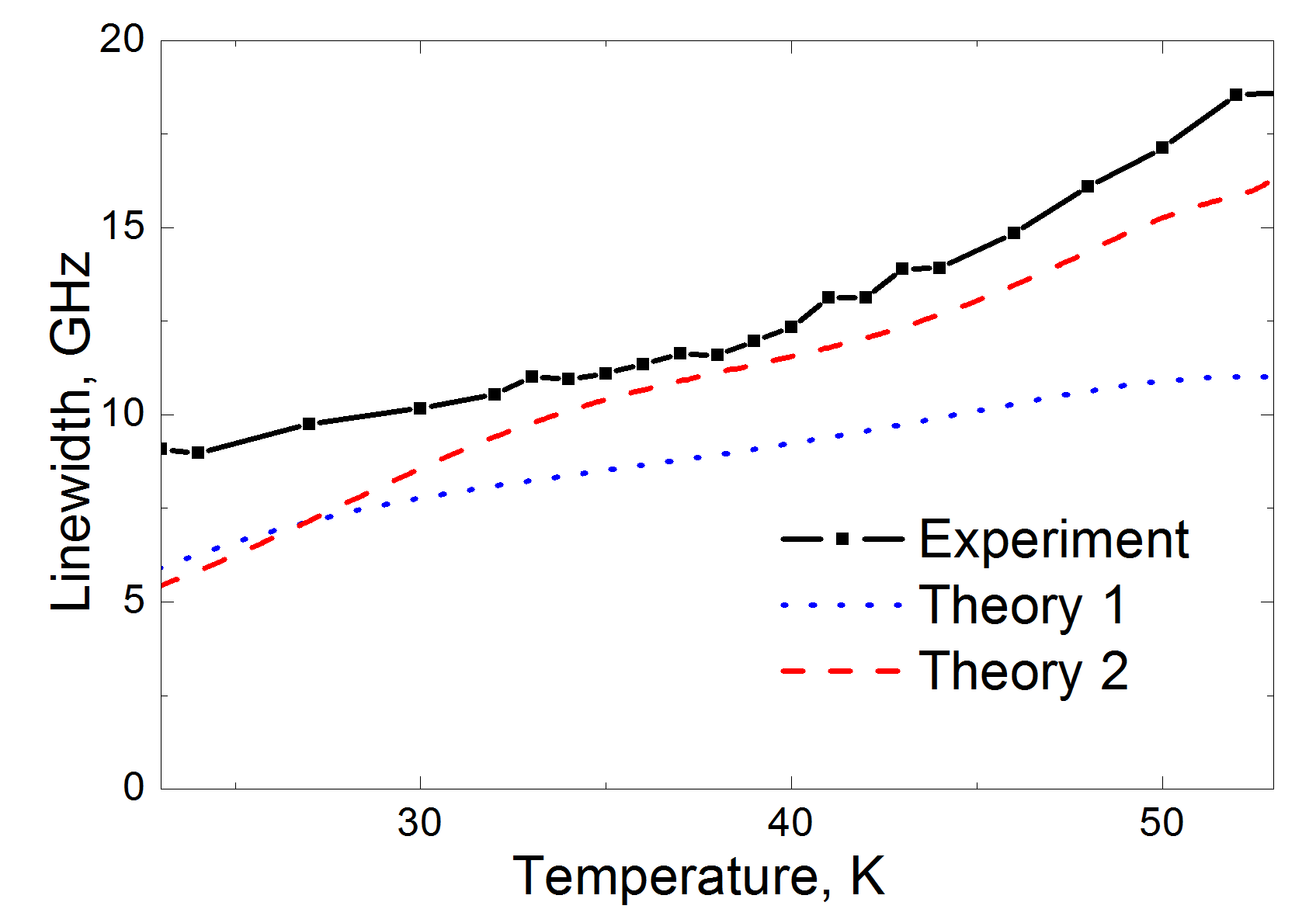}\par
\hspace{4.3cm}\textit{(a)}\hspace{8.2cm}\textit{(b)}
\captionof{figure}{(\textit{a}) Experimental  critical current $I_{c}$ as a function of temperature $T$ (solid black line) fitted by de Gennes proximity effect model (dashed red line); (\textit{b})  Generation linewidth  a function of temperature (black solid line). Dashed red and blue dotted lines correspond to the two theoretical models based on formula (\ref{linewidth}).}
\label{fig3}
\vspace{-0.5cm}
\end{figure}

Experimental dependence of the junction critical current on temperature is shown in figure \ref{fig3}(\textit{a}). To describe the results we assume
a  linear dependence of the normal segment length with temperature $T$ and bias current $I$:
\begin{equation}
l(I,T)=l_{0}(T)+\alpha\cdot I
\end{equation}
\begin{equation}
l_{0}(T)=l_{max}\left(1+\frac{T-T_{c}}{T_{c}-T_{c}^{\prime}}\right)\hspace{1cm} T_{c}^{\prime}<T<T_{c}
\end{equation}
where $\alpha$ is a fitting parameter and $l_{max}$ is the maximum length of the normal segment in the absence of bias current at  $T=T_{c}$. $T_{c}^{\prime}$ is a temperature at which the entire damaged area becomes superconducting ($l(T_{c}^{\prime}$)=0). We can estimate $T_{c}^{\prime}\sim$ 20K  and $l_{max}\sim$ 100nm (i.e. much larger than the slit width, see e.g. \cite{kahlmann1998}) from the R(T) curve presented on the figure \ref{fig1}(\textit{b}) and Monte-Carlo (TRIM) simulation. \par
For conventional s-wave superconductors, $I_{c}(T)$ characteristics were extensively investigated for different types of junctions (see e.g. review \cite{golubov2004}) including $SS^{\prime}S$-structures with $T_{c}\neq T_{c}^\prime$ (\cite{kupriyanov1981,kupriyanov1982}). However, the applicability of these models for d-wave superconductors is not straightforward. Here, we use a simple proximity effect approximation as described by the de Gennes model:     
\begin{equation}
I_{c}(I,T)=I_{0}(I,T)\cdot \left(1-\frac{T}{T_{c}}\right)^2\cdot \frac{\kappa(I,T) l(I,T)}{sinh(\kappa(I,T) l(I,T))}
\label{Ic}
\end{equation}
where $\kappa(T)$ is the reversed decay length given by
\begin{equation}
\kappa(T)=\left(\frac{D\hbar}{2\pi k_BT}\right)^{-1/2}\cdot \left(1+\frac{2}{ln(T/T_{c}^{\prime \mathrm{eff}})}\right)^{-1/2}.
\label{kappa}
\end{equation}
Here, $D$ is the electron diffusion constant, $k_B$ is the Boltzmann constant and  $T_{c}^{\prime \mathrm{eff}}$ is an effective critical temperature corresponding to an average of $T_{c}^{\prime} $ along the junction. $I_{0}(I,T)$ is the $T$=0 value of the critical current which is given by  the expression :
\begin{equation}
I_{0}(I,T)=\frac{\pi \Delta_{0}^2}{4eR_{N}(I,T)k_{B}T_{c}}
\label{I0}
\end{equation}
 where $\Delta_{0}$ is the YBa$_2$Cu$_3$O$_7$ energy gap and $R_N$ the normal resistance of the junction.
The experimental critical current has to be compared with the one extracted for each particular temperature as a solution of the equation $I_{c}(I,T)=I$. Although the approach is expected to be strictly accurate only in the high temperature limit, a good agreement is obtained in the whole Josephson regime (see red dashed line on the figure \ref{fig3}(\textit{a})).  
\subsection{Generation linewidth }
To analyze the experimental generation linewidth we describe the junction dynamics within the simple RSJ model and the Likharev-Semenov model \cite{likharev1986dynamics} for nonlinear noise power transformation from high to low frequency. For simplicity we only take into consideration the thermal noise source with a current spectral density\footnote[1]{We define spectral density for both negative and positive frequencies through expression $\langle \left.I_{fluct}^2\rangle\right|_{d\omega}=2S_I(\omega)d\omega$ for time averaged value of squared fluctuation current.} $S_{I}(0)=\frac{k_{B}T}{\pi R_{N}}$. The voltage spectral density can then be written as $S _{V}(0)=S^\prime_{I}(0)\cdot R_{d}^{2}$, where $R_d$ is the differential resistance of the junction and $S^\prime_{I}(0)=S_{I}(0)(1+\frac{I_{c}^2}{2I^{2}})$. In the frame of the Likharev-Semenov model the generation linewidth can be expressed as
\begin{equation}
\Delta f=\frac{4\pi}{\Phi_{0}^2}\cdot \frac{k_{B}TR_{d}^2}{R_{N}}\cdot \left(1+\frac{I_{c}^2}{2I^2}\right)\hspace{0.3cm}[Hz]
\label{linewidth}
\end{equation}
where $\Phi_{0}$ is the magnetic flux quantum.\par
As a first order approximation we can directly apply formula (\ref{linewidth}) using  the experimental normal resistance $R_N=V/I$ and the differential resistance $R_{d}=dV/dI$ of the junction at the working point.  The result of this simple estimation is shown in figure \ref{fig3}(\textit{b}) with blue dots-line and referenced as "theory 1". A qualitative agreement is obtained with the experimental linewidth (black solid line) including  in the low temperature range where the junction approaches a flux-flow regime. However, "theory 1" demonstrates a slower increase with temperature than the experimental one in the high temperature range area where the RSJ-model is expected to be more accurate.\par
To refine our analysis we propose another approach that explicitly takes into account the temperature and the current dependence of $R_{N}$.  Following equations (1) and (2), we assume a linear dependence
\begin{equation}
R_{N}(I,T)=R_{0}(T)+\beta\cdot I
\label{RN}
\end{equation}
\begin{equation}
R_{0}(T)=R_{max}\left(1+\frac{T-T_{c}}{T_{c}-T_{c}^{\prime}}\right)\hspace{1cm} T_{c}^{\prime}<T<T_{c}
\label{R0}
\end{equation}
where $R_{0}(T)$ is the normal segment resistance in the absence of bias current and $R_{max}$ is the value of the resistance at $T=T_{c}$ (see figure \ref{fig1}(\textit{b})). \par
According to the simple RSJ-model, when the junction is biased with a dc current $I$ the voltage across the junction is $V=R_{N}\sqrt{I^{2}-I_{c}^{2}}$ and correspondingly the differential resistance is $R_{d}=\frac{dV}{dI}=\frac{R_{N}I}{\sqrt{I^{2}-I_{c}^{2}}}$. Introducing equations (\ref{RN}) and (\ref{R0}) in the previous expressions of the voltage and the differential resistance allows calculating the generation linewidth (\ref{linewidth}) with only one fitting parameter $\beta$. The fit presented on  figure \ref{fig3}(\textit{b}) (red dashed line) and referenced as "theory 2" shows a quantitative agreement of the model with the experimental data. In particular, the temperature dependence of the generation linewidth is correctly reproduced. This remarkable behavior indicates that the noise of the junction is mainly limited by the broadband thermal noise and that additional sources of noise can be neglected at first order.  Small discrepancies between experiment and theoretical model could originate from external noise absorbed by the unshielded junction in the refrigerator.\par
In this analysis, we used the Likharev-Semenov model which has been proposed for short Josephson junction corresponding to the regime $L<\lambda_J$ where $\lambda_{J}=\sqrt{\frac{\Phi_{0}}{2\pi\mu_{0}j_{c}d}}$ is the Josephson penetration length ($\lambda_{L}$ is the London penetration length, $d=2\lambda_{L}+l$, $j_{c}$ is the critical current density and $\mu_{0}$ is the vacuum permeability). Assuming a London penetration depth $\lambda_{L}(0)$ $\sim$ 200~nm  for our YBa$_2$Cu$_3$O$_7$ junctions  \cite{wolf2013}, the Josephson length is $\lambda_{J}\approx600$~nm for  a critical current of 100 $\mu$A which is comparable to the width of the junction $L$=750 nm. Therefore, for the lower part of the temperature range investigated here, the junction is not strictly short and the Likharev-Semenov model may not give a very accurate result. To our knowledge there is no commonly accepted general theory on generation linewidth for long Josephson junctions, although known results (\cite{golubov1996}, \cite{microwave_superconductivity}) suggest that it is somewhat broader compared to the short junction case.
\section{Conclusion}
In conclusion, we have measured the transport characteristics as well as the Josephson linewidth dependence with temperature for ion-irradiated YBa$_2$Cu$_3$O$_7$ junctions. A good agreement is obtained with the Likharev-Semenov model which indicates that the linewidth of the junction is mainly limited by the broadband thermal noise. As already mentioned, a long term goal in the field is to build a frequency-tunable oscillator by synchronizing a large number of Josephson junctions.  In previous experiments (\cite{malnou2012,malnou2014}), we demonstrated that ion-irradiated junctions could be operated as broadband Josephson mixers up to $\sim$ 400 GHz and possibly higher frequencies provided a better impedance matching with the antenna and the CPW line. According to the experimental data shown in figure \ref{fig3}(\textit{b}), the ion-irradiated junctions display a typical Josephson linewidth of approximately 10 GHz at 40K. Synchronizing an array of 10$\times$10 junctions should therefore allow generating a coherent $\simeq$ 100 MHz linewidth oscillation in the lower part of the THz spectrum. Combined with a Josephson mixer such arrays could be used as a frequency-tunable local oscillator to build a fully integrated detector.
\section*{Acknowledgments}
The authors thank Yann Legall for ion irradiations. This work has been supported by the T-SUN ANR ASTRID program, the Emergence program Contract of Ville de Paris and by the R\'{e}gion Ile-de-France in the framework of the DIM Nano-K and Sesame programs.
\bibliographystyle{ieeetr} 
\bibliography{linewidth_paper}
\end{document}